\documentclass[aps,pra,twocolumn,amsmath,amssymb,nofootinbib,showpacs,superscriptaddress]{revtex4-1}

\usepackage{graphicx}
\usepackage{dcolumn}
\usepackage{bm}
\usepackage{braket}
\usepackage{comment}
\usepackage{latexsym}
\usepackage{graphics}
\usepackage{graphicx}
\usepackage{epsfig}
\usepackage{color}
\usepackage{bm}
\usepackage{amsmath}
\usepackage{amssymb}
\usepackage{amsthm}
\usepackage{dcolumn}
\usepackage{float}
\usepackage{hyperref}
\usepackage{color}
\usepackage{epstopdf}
\usepackage{cleveref}
\usepackage{soul}
\usepackage{url}
\usepackage{tabularx}
\usepackage{array}
\usepackage{enumitem}

\usepackage[svgnames]{xcolor}
\hypersetup{hidelinks,colorlinks=true,allcolors=DarkBlue}

\usepackage{diagbox}

\newcommand{\var}{\ensuremath{s}}
\newcommand{\bx}{\ensuremath{\mathbf{x}}}
\newcommand{\bs}{\ensuremath{\mathbf{s}}}
\newcommand{\be}{\ensuremath{\mathbf{e}}}

\newcommand{\bof}{\ensuremath{\mathbf{f}}}
\newcommand{\reals}{\ensuremath{\mathbb{R}}}

\newcommand{\sign}{\operatorname{sign}}

\usepackage{mathtools}
\DeclarePairedDelimiter\ceil{\lceil}{\rceil}

\newcommand{\red}[1]{#1}

\usepackage[normalem]{ulem}

\newtheorem{problem}{Problem}

\theoremstyle{remark}

\theoremstyle{definition}

\theoremstyle{plain}

\begin{document}
	
	\title{Polynomial unconstrained binary optimization inspired by optical simulation}%
	
	
	\author{D.A. Chermoshentsev}
	\email{e-mail: dmitriy.chermoshentsev@phystech.edu}
	\affiliation{Russian Quantum Center, Skolkovo, Moscow 143025, Russia}
	\affiliation{Moscow Institute of Physics and Technology, Moscow Region 141701, Russia}
	\affiliation{Skolkovo Institute of Science and Technology, Skolkovo, Moscow, 143025, Russia}
	\affiliation{QBoard, Skolkovo, Moscow 143025, Russia}
	\author{A.O. Malyshev}
	\affiliation{Clarendon Laboratory, University of Oxford, Parks Road, Oxford OX1 3PU, UK}
	\author{M. Esencan}
	\affiliation{Clarendon Laboratory, University of Oxford, Parks Road, Oxford OX1 3PU, UK}
	\author{E.S. Tiunov}
	\affiliation{Russian Quantum Center, Skolkovo, Moscow 143025, Russia}
	\affiliation{Moscow Institute of Physics and Technology, Moscow Region 141701, Russia}
	\author{D. Mendoza}
	\affiliation{Department of Chemistry, University of Toronto, 80 St. George Street, Toronto, ON M5S 3H6, Canada}
	\affiliation{Department of Chemistry and Chemical Biology, Harvard University, 12 Oxford Street, Cambridge, MA 02138, USA}
	\author{A. Aspuru-Guzik}
	\affiliation{Department of Chemistry, University of Toronto, 80 St. George Street, Toronto, ON M5S 3H6, Canada}
	\affiliation{Canadian Institute for Advanced Research (CIFAR) Senior Fellow, Toronto, ON M5S 1M1, Canada}
	\affiliation{Canada CIFAR AI Chair Lebovic Fellow, Vector Institute, Toronto, ON M5S 1M1, Canada}
	\affiliation{Department of Computer Science, University of Toronto, 40 St. George Street, Toronto, ON M5S 2E4, Canada}
	\author{A.K. Fedorov}
	\affiliation{Russian Quantum Center, Skolkovo, Moscow 143025, Russia}
	\affiliation{QBoard, Skolkovo, Moscow 143025, Russia}
	\affiliation{National University of Science and Technology ``MISIS'', Moscow 119049, Russia}
	\author{A.I. Lvovsky}
	\affiliation{Russian Quantum Center, Skolkovo, Moscow 143025, Russia}
	\affiliation{Clarendon Laboratory, University of Oxford, Parks Road, Oxford OX1 3PU, UK}
	
	\date{\today}
	
	\begin{abstract}
		We propose an algorithm inspired by optical coherent Ising machines to solve the problem of polynomial unconstrained binary optimization (PUBO).  
        We benchmark the proposed algorithm against existing PUBO algorithms on the extended Sherrington-Kirkpatrick model and random third-degree polynomial pseudo-Boolean functions, and observe its superior performance. 
        We also address instances of practically relevant computational problems such as protein folding and electronic structure calculations with problem sizes not accessible to existing quantum annealing devices. 
        The application of our algorithm to protein folding and quantum chemistry problems sheds light on the shortcomings of  approximating the electronic structure problem by a PUBO problem, which, in turn, puts into question the applicability of the unconstrained binary optimization formulation, such as that of quantum annealers and coherent Ising machines, in this context. 
		
	\end{abstract}
	
	\maketitle
	
	\section{Introduction}\label{sec:intro}
	
	A notable example of a combinatorial optimization problem is the problem of polynomial unconstrained binary/spin optimization (PUBO/PUSO), which is formulated as follows.
	\begin{problem}\label{problem:pubo}
		Find the global minimum point of a pseudo-Boolean ``energy" function $H_k: \bs=\{-1, +1\}^N  \rightarrow \reals{}$ such that: 
		\begin{equation}\label{equ:pubo}
			H_k(\bs) =  \sum_{i_1}J_{i_1}^{(1)} \var_{i_1} + \ldots+ \sum_{i_{1} <\ldots< i_{k}} J_{i_{1}\ldots i_{k}}^{(k)}\var_{i_{1}}\ldots \var_{i_{k}}.
		\end{equation}
		Here $N$ is the dimension of a problem instance (problem size), $k$ is the function degree, the summations run from 1 to $N$, and the coupling tensors $J^{(\cdot)}$ with real coefficients are given.
	\end{problem}
	Historically the dominant subject of the PUBO-related research was its particular case ($k=2$), also known as the \emph{quadratic unconstrained binary optimization (QUBO)} problem.
	Among naturally arising QUBO problems of practical value are finding the ground state of a spin-glass~\cite{Wu1991} and the problem of constrained via minimisation in very-large-scale integration design~\cite{comb_opt_book_2}.
	The interest to the QUBO problem is also motivated by the fact that any PUBO problem can be reduced to a QUBO problem by increasing the problem dimension~\cite{Biamonte2008}.
	In addition, the NP-complete problem MAX-CUT can be reduced to a QUBO problem instance~\cite{comb_opt_book_2, karp_np_problems}. Therefore, if one has an efficient procedure to solve the QUBO problem, one may efficiently solve any other problem in the complexity class \textsc{NP}~\cite{karp_np_problems}.
	
	Traditionally, QUBO instances were addressed by means of ordinary combinatorial optimization methods, which include the branch-and-cut method~\cite{max_cut_branch_and_cut}, semi-definite programming~\cite{max_cut_semi_definite_1, max_cut_semi_definite_2}, polynomial time approximation schemes~\cite{max_cut_ptas}, simulated annealing~\cite{max_cut_simulated_annealing}, etc.
	An alternative perspective to solve QUBO problems is to build a special-purpose computing device, and this approach has been widely investigated over the last decades~\cite{Boixo2013, yamaoka201520k, puri2017quantum, tsukamoto2017accelerator, PhysRevLett.122.213902, Pierangeli:20, Roques-Carmes2020Jan}, see Vadlamani {\it et al.} \cite{doi:10.1073/pnas.2015192117} for a review.
	The idea is to relate a second degree pseudo-Boolean function to the energy landscape of some physical system governed by the Ising Hamiltonian,
	and to let the system evolve into its ground state.
	
	A special family of special-purpose devices is the class of D-Wave quantum annealers, which use superconducting qubits coupled by magnetic fields as the underlying platform \cite{Boixo2013}.
	D-Wave annealers were used for a number of combinatorial optimization tasks ranging from logistics to quantum chemistry and material simulation \cite{Neukart2017,Streif2019, Li2018, Perdomo-Ortiz2012, Boev2020,  Amin2018,  Biamonte2017, Venturelli2015, Adachi2015}. 
	A downside of these processors is limited connectivity between qubits: a typical state-of-the-art D-Wave Advantage processor has 5000 physical qubits, but each is only coupled to 15 others~\cite{d_wave_advantage}. 
	The ability of D-Wave processors to demonstrate quantum computation speed-up is a subject of ongoing research~\cite{no_d_wave_speedup}.
	
	Another special-purpose devices are coherent Ising machines (CIM), which store information about optimization variables in optical pulses and use an optoelectronic feedback loop to implement the couplings between them~\cite{Wang2013,Mcmahon2016,Inagaki2016}. CIMs have no restrictions on the connections between variables and are capable of dealing with QUBO functions which act on up to 2048 variables and are ``dense'', i.e. have $~\Omega(N^2)$ non-zero terms~\cite{Inagaki2016}. 
	Despite the fact that quantum properties of these devices are also subject to debate, it was recently shown that CIMs significantly outperform D-Wave processors in dealing with dense QUBO functions~ \cite{Hamerly2019}, although this claim was disputed by the D-Wave team \cite{McGeoch2018}.
	
	An endeavour to overcome disadvantages of existing special-purpose devices resulted in the development of so-called \textit{quantum-inspired} optimization algorithms~\cite{Arrazola2019, Goto2019, Goto2021, Tatsumura2021}.
	For example, in Ref.~\cite{Tiunov2019}, some of us proposed a ``SimCIM" algorithm to solve QUBO problems by classicaly simulating the optimization procedure of CIMs. 
	SimCIM can be implemented using GPU-accelerated matrix-vector multiplications and achieves solution speed and quality that is comparable to that of CIM and higher than many competing software algorithms.
	
	In~\cite{Stroev2021} authors make first steps towards quantum-inspired polynomial optimization --- they propose an algorithm adopted from operation of networks of nonequilibrium polariton condensates. 
	In this paper we advance further along this direction and introduce \emph{PolySimCIM} --- a generalisation of  SimCIM~\cite{Tiunov2019} that enables direct solving of PUBO problems of degrees $k>2$. 
	This extends the application domain of our algorithm to problems initially formulated as PUBO, such as finding the tertiary structure of proteins,  calculating the electronic structure of  molecules, and solving box constrained Diophantine equations \cite{Perdomo-Ortiz2019, Babbush2013,  quantum_ising_to_classical, diophantine_equations}.
	While any PUBO problem can in principle be reduced to QUBO, this reduction is associated with an overhead which in some cases can lead to an increase of the problem size $N$ by several orders of magnitude, as we demonstrate in Sec.~\ref{app:qubo_to_pubo}. PolySimCIM thus helps avoiding this overhead. 
	We benchmark our algorithm on synthetic data of random third-degree pseudo-Boolean functions and reveal its advantage compared to other state-of-the-art optimization algorithms (including the one presented in~\cite{Stroev2021}). 
	Additionally, we test its performance on the protein folding and electronic structure problems~\cite{Babbush2014, quantum_ising_to_classical} and demonstrate that the proposed approach allows dealing with problem sizes previously unreachable for special-purpose QUBO devices (in particular, the D-Wave machine) or their simulators.
	

	\section{SimCIM for polynomial binary optimization}\label{sec:poly_simcim}
	
	We begin by recapping the functionality of SimCIM~\cite{Tiunov2019}. As mentioned, this algorithm \emph{sim}ulates \emph{CIM}s, which are special-purpose devices tailored to solve the QUBO problem. 
	Physically each CIM consists of an optical parametric oscillator  formed by degenerate parametric amplifier (single-mode squeezer) inside a  fiber loop, with  $N$ optical pulses travelling in that loop.
	In each roundtrip, each pulse is subjected to (i) measurement of the position quadrature  and (ii) phase-space displacement along the position axis.
	The displacement is computed from the measured position quadratures of other pulses and the coefficients of the function $H_2(\bs)$ that is being optimised. After multiple roundtrips, each pulse is amplified to a coherent state of certain intensity. 
	Thanks to the phase-sensitive nature of the parametric amplification, the phase $\varphi_i$ of each pulse stabilizes at either to $0$ or to $\pi$. 
	The sequence of these phases is used to obtain the sequence of $\var_i=\pm 1$ corresponding to the solution produced by CIM \eqref{equ:pubo}.
	
	To fully simulate a CIM, one should consider evolution of both complex position and momentum quadratures as well as linear and nonlinear losses present in the system. However, SimCIM  restricts analysis to only the real part of the position quadrature and simplifies the effect of noise and nonlinear losses.
	Each iteration consists of two steps as follows.
	
	{\it Step 1.} Calculate the displacement and apply it to the vector of continuous quadratures (set to zero before the start of simulation):
	\begin{equation}\label{equ:displacement}
		\Delta x_{i_1}(t) = \nu(t) x_{i_1} + \xi \left(\sum_{i_2} J_{{i_1}{i_2}}^{(2)}x_{i_2} + J_{i_1}^{(1)}\right) + f_{i_1}(t)
	\end{equation}
	Here $\nu(t)$ accounts for combination of (time-dependent) pump amplitude and linear loss, $\xi$ is equivalent to learning rate in the conventional gradient descent algorithm,
	and $f_i$ is Gaussian noise with variance $\sigma^2$. To accelerate the convergence, we use the momentum method \cite{Qian1999}
    with the momentum parameter of $\alpha$ = 0.99.
	
	{\it Step 2.} Apply an activation threshold function to each variable to account for the saturation:
	\begin{equation}\label{equ:saturation}
		x_{i_1}=\begin{cases} x_{i_1} & \mbox{for}\ |x_{i_1}| < x_{\rm sat}, \\
			x_{\rm sat} &\mbox{otherwise}.
		\end{cases}
	\end{equation}
	In the above equations, $\nu(t)$, $\xi$, $\sigma$, and $x_{\rm sat}$ 
	are hyperparameters of the algorithm and require fine-tuning for each problem instance. In the original paper,  $\nu(t)$ grows with time according to the hyperbolic tangent law. After the last iteration, the achieved ``solution'' is evaluated as $\var_i := \sign{x_i}$.

	
	The second term in Eq.~\eqref{equ:displacement} 
	is basically a partial derivative with respect to $x_{i_1}$ of a continuous function $H_k$(\bx) obtained from $H_k(\bs)$ by extending the domain of the latter from $\{-1, 1\}^N$ to $\reals^N$ (here $k=2$). 
	Hence, Step 1 can be understood as calculating the displacement vector as
	\begin{equation}\label{equ:vectorised_displacement}
		\Delta \bx(t) = \nu(t) \cdot \bx + \xi \cdot \nabla H_k(\bx) + \bof(t).
	\end{equation}
	with 
	$$\nabla H_k(\bx)=\frac{\partial H_k(\bx)}{\partial x_{i_1}}=\sum_{i_2} J_{{i_1}{i_2}}^{(2)}x_{i_2} + J_{i_1}^{(1)}$$ for the SimCIM case.

\vspace{1em}
Similarly to SimCIM, PolySimCIM follows the gradient of the energy function (of an arbitrary degree) supported with some noise --- so that the algorithm can get out of local optima. More specifically, Step 1 of SimCIM is replaced in PolySimCIM by 

{\it Step 1$'$.} Calculate the displacement and apply it to the vector of quadratures according to 
Eq.~\eqref{equ:vectorised_displacement} with
	\begin{equation}\label{equ:polynomial_gradient}
		\nabla H_k(\bx) = J_{i_{1}}^{(1)} +\ldots+ \sum_{i_{2}<\ldots <i_{k}} J_{i_{1}\ldots i_{k}}^{(k)}x_{i_{2}}\ldots x_{i_{k}}.
	\end{equation}
	
The annealing parameter $\nu(t)$ used in calculations is a shifted hyperbolic tangent which is determined by three hyperparameters $O, D, S$. The extended form of $\nu(t)$ is as follows:
\begin{equation}
    \nu(t) = O \cdot \tanh\left(S \left(\frac{t}{N} - 0.5\right)\right) - D,
\end{equation}
where $O$, $D$, $S$ are hyperparameters, $N$ is the number of steps in a single run.
	
	

	\section{Benchmarking}\label{sec:benchmarking} 
	\subsection{Comparison with PUBO algorithms}
		We begin by benchmarking PolySimCIM against existing optimization algorithms on pseudo-Boolean functions of third degree:
		\begin{equation}\label{supersymm_ising}
		H_3(\bs) = \sum_{i_{1} < i_{2} < i_{3}} J_{i_{1}i_{2}i_{3}}\var_{i_{1}} \var_{i_{2}} \var_{i_{3}}.
	\end{equation}
	We consider three classes of such functions based on the structure of tensor $J_{i_{1}i_{2}i_{3}}$.
	Functions in Class I are those with each element $J_{i_1 i_2 i_3}$ randomly set to $1$ or $-1$ with probability $0.5$. In a sense, they are generalisation of the paradigmatic Sherrington-Kirkpatrick model~\cite{Fu_1986} to the case of third degree pseudo-Boolean functions. 
	Functions in Class II also have dense tensors, but each element $J_{i_1 i_2 i_3}$ is drawn uniformly from $[-1, 1]$ interval.
	Class III includes pseudo-Boolean functions with random \emph{sparse} tensors, which are in practice obtained by taking a function from Class II and setting each tensor element to zero (i.e. ``dropping'' it) with probability $0.9$. 
	We considered problem sizes $N$ ranging from 10 to 100 in steps of 10.
	For each problem dimension, we generated ten pseudo-Boolean functions.
		
	\begin{figure*}
		\centering
		\includegraphics[width=\textwidth]{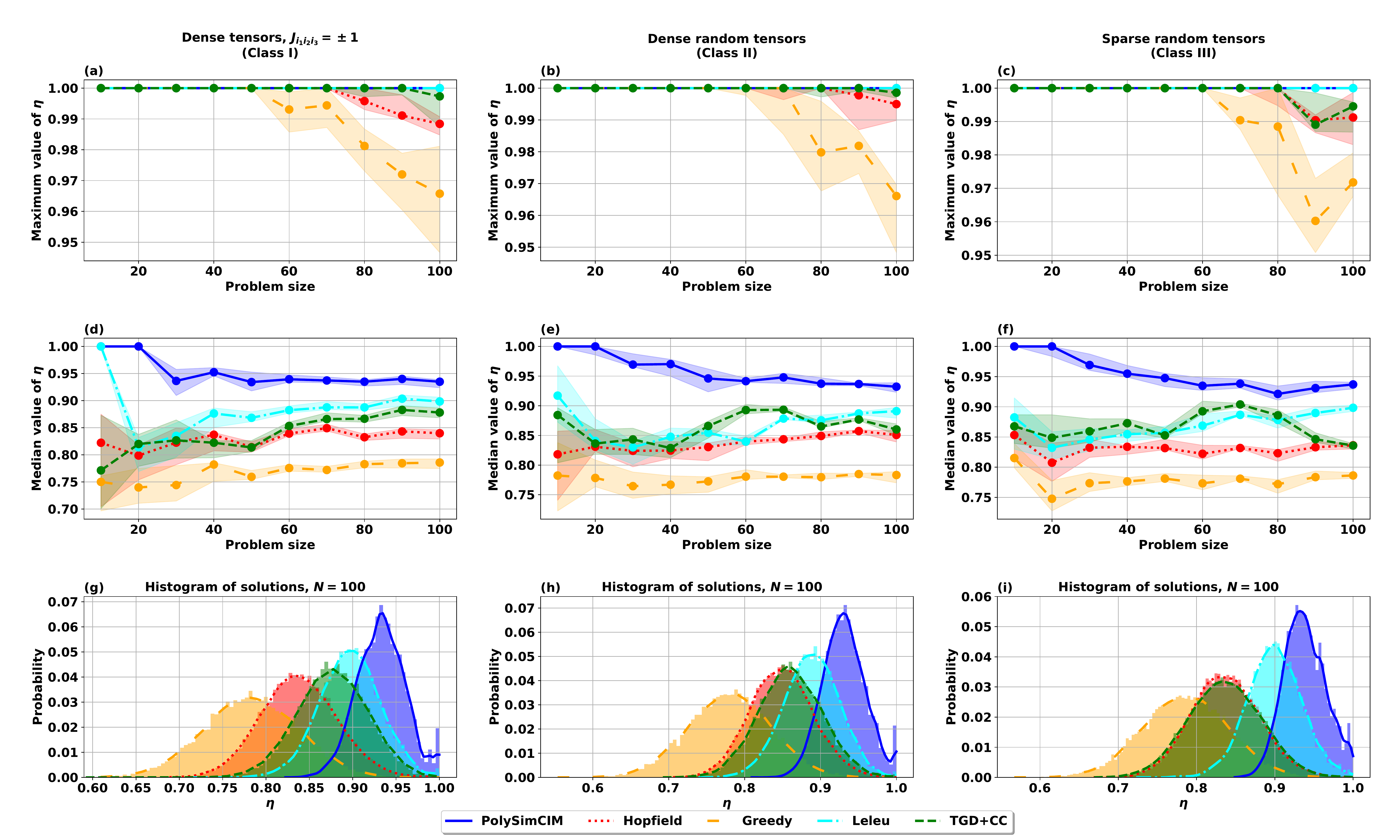}
		\caption{Performance comparison of polynomial optimization algorithms on the generalised Sherrington-Kirckpatrick, dense random and sparse random tensors. Top two rows: maximum (a–c) and median (d–f) achieved figure of merit $\eta$  over 2000 runs of each algorithm for 10 problem instances corresponding to a specific problem class and size. Lines and shaded regions show, respectively, the medians and the 25-75 percentiles of these maxima and medians over each set of problem instances. Bottom row (g–i): histograms with kernel density estimation (KDE)} of $\eta$ for each algorithm aggregated over all runs and problem instances of size $N = 100$.
		\label{fig:benchmarking}
	\end{figure*}
	
	We compare PolySimCIM with four optimization algorithms, which are also based on iterative updates of variables (Table~\ref{tab:solvers_reference}).
	The first three of them use continuous representation of optimization variables and update them based on the energy function gradient. 
	The first algorithm adapts higher-order Hopfield-Tank neural networks and is described in Ref.~\cite{Joya2002}. 
	The second approach (Leleu) generalises the method of Ref.~\cite{Leleu2019} to $k>2$: it uses an additional vector of time-dependent ``error variables"  $\be$ that helps escaping local minima. We used the realisation of the algorithm described in section ``Benchmark results on the G-SET" in Ref.~\cite{Leleu2019}, straightforwardly adapting the injection term to the polynomial case.
    The third algorithm (TGD+CC) \cite{Stroev2021} permits the amplitude vector $\bx$ to take on complex values and makes the gain parameter $\nu(t)$ dependent on these amplitudes.  
    Additionally, whenever the algorithm appears to reach a steady state, it adds complex parts to a small number of randomly chosen elements of $J_{i_{1}i_{2}i_{3}}$, and removes them when the system leaves the local minimum.
	The fourth algorithm is discrete greedy search: at each iteration it moves to the neighbouring vertex of $\{-1, 1\}^N$ hypercube which has the smallest value of $H_k(\bs)$. 
	
	\begin{table}
	\centering
	\begin{tabular}{|m{2.5cm}|m{5cm}|}
		\hline
		{\bf Algorithm} & {\bf Variable evolution law} \\
		
		\hline
		PolySimCIM &  $\Delta \bx(t) = \nu(t) \cdot \bx + \xi \cdot \nabla H_k(\bx) + \bof(t)$\\
		\hline
		Hopfield-Tank & $\Delta \bx(t) = -\bx + \xi \cdot \nabla H_k(\tanh(\bx/\beta)) $\\
		\hline
		Leleu & $\Delta \bx(t) = \nu(t) \cdot \bx -  \bx^3 + \xi \be(t)\nabla H_k(\bx)$\\
		&$\Delta \be(t) = -\beta(t)(\bx^2 -a(t)) \be$\\
		\hline
		TGD+CC&$\Delta x_{i_1}(t)=\xi\big( x_{i_1}(\nu_{i_1}(t)-| x_{i_1}|^2)+ \sum \limits_{i_{2}<\ldots<i_{k}}^{N} J^{(k)}_{i_1 i_2\ldots  i_k} x_{i_{2}} x_{i_{3}}\ldots x^{*}_{i_{k}}\big)$
		$\Delta \nu_{i_1}(t) = \xi \varepsilon(\rho_{\rm th} - | x_{i_1}|^2)$\\
		\hline
		Greedy&${\bs}(t+1)=\arg \displaystyle \min_{\bs:|\bs-\bs(t)|=2}H_k(\bs)$\\
		\hline
	\end{tabular}
	\caption{Update rules for the considered polynomial optimization algorithms. All arithmetic operations (i.e. addition, multiplication, exponentiation etc.) are element-wise. All variables in equations except for evolving vectors and pseudo-Boolean functions are hyperparameters of the corresponding algorithms. After each iteration step of the PolySimCIM the activation threshold function determined by the Eq.~\eqref{equ:saturation} is applied to the amplitudes.}
	\label{tab:solvers_reference}
	\end{table}

	For each problem instance, 
	we performed 2000 simultaneous runs of each algorithm. The hyperparameters of PolySimCIM, Leleu, Hopfield-Tank, and TGD+CC algorithms were optimized with a machine-learning online optimization package M-LOOP \cite{Wigley2016}  (see Appendix \ref{app:hyperparameter_search} for details).     
    The realisation of TGD+CC was provided by the authors of Ref.~\cite{Stroev2021}. The number of steps was equal to 1000 for each method. 
	
	A set of hyperparameters was found for each class of energy functions on a single problem instance of each dimension and then used for all other problems of the same class and dimension. 
	The performance of each algorithm could be further optimized by picking a set of hyperparameters for each problem instance individually. Even deeper optimization could be reached by choosing the type of nonlinearity for a given algorithm and problem instance \cite{Bohm2021}. However, such in-depth individual optimization would slow down the process if multiple problem instances of similar types are to be solved, so we choose not to use this strategy.

	The starting configurations of the Hopfield-Tank and Leleu algorithms were randomly initialised from a normal distribution $\mathcal{N}(0, 10^{-4})$, and for the Greedy algorithm the starting  configurations were chosen randomly and uniformly from $\{-1, 1\}^N$. 
	Different runs of the \red{Hopfield-Tank, Leleu and Greedy} algorithms produced different results due to these random initial conditions. 
	For the PolySimCIM and TGD+CC algorithms, different results across runs were obtained due to the random noise present in their update routines.
	Each run of each algorithm produced an estimate $\hat{\bs}_{\rm min}$ for the optimal spin configuration.  
	Next, we picked the minimum among all $5 \cdot 2000 = 10 000$ obtained values of $\hat{\bs}_{\rm min}$ (i.e. among the results produced by all algorithms), which we denote as $\bs_{\rm min}$.
	As a figure of merit for each run we use the following ratio:
	\begin{equation*}
		\eta = \frac{H_3(\hat{\bs}_{\rm min})}{H_3(\bs_{\rm min})}.
	\end{equation*}
	The energy functions can take both positive and negative values and so does $\eta$.
	However, in all our experiments $H_{3}(\bs_{\rm min}) < 0$, which implies that $\eta \in (-\infty, 1]$, and therefore higher solution quality corresponds to $\eta$ being closer to unity.

	To compare the performance, we aggregate the information about values of $ \eta $ achieved in different runs in three possible ways, as depicted in Fig.~\ref{fig:benchmarking}. 
	First, for each problem instance, we pick the best value of $\eta$ achieved by the given algorithm among all runs on a given problem instance and we take the median of these values among all problem instances of a given problem size [Fig.~\ref{fig:benchmarking}(a--c)].  
	Second, we find the median of $\eta$'s achieved by the given algorithm among all runs on a  given problem instance, and then plot the medians of these medians over the 10 problem instances for each problem size [Fig.~\ref{fig:benchmarking}(d--f)].
	Finally, we plot the histograms  of the $\eta$ values obtained for all problems of size $N=100$ 
	with each algorithm 
	[Fig.~\ref{fig:benchmarking}(g--i)].
	
	Figures \ref{fig:benchmarking}(a--c) show that the Hopfield-Tank and Greedy algorithms perform worse than others: they systematically fail to achieve the best known solutions for large problem sizes, except for the Class I pseudo-Booleans, for which the Greedy algorithm successfully reaches $\eta = 1$ [Fig.~\ref{fig:benchmarking}(a--c)].
	In contrast, PolySimCIM manages to achieve $\eta = 1$ for each problem instance of each problem size and for all classes of pseudo-Boolean functions. Furthermore, as evident from the histograms in Fig.~\ref{fig:benchmarking}(g--i), the quality of  PolySimCIM solutions is consistently high for different runs of the algorithm and different problem instances. In this latter respect, PolySimCIM surpasses Leleu, which is also able to find the optimal solution in most cases, but shows higher variances among the runs and problem instances.

	\begin{table}
	\centering
	\begin{tabular}{|m{3cm}|m{3cm}|}
		\hline
		{\bf Algorithm} & {\bf Execution time} \\
		
		\hline
		PolySimCIM &  2.7 s\\
		\hline
		Hopfield-Tank & 1.2 s \\
		\hline
		Leleu & 4.4 s\\
		\hline
		TGD+CC& 24.3 s\\
		\hline
		Greedy& 8522.8 s\\
		\hline
	\end{tabular}
	\caption{Average execution time of the algorithms for tensors with dimension $N=100$ on the  NVIDIA 2080Ti GPU.
	PolySimCIM, Hopfield-Tank, Leleu make 1000 iterations. TGD+CC makes 1000 iterations before complex coupling switching and 1000 iterations after it. Each algorithm made 2000 simultaneous runs.}
	\label{tab:solvers_reference_times}
	\end{table}
	
\subsection{Comparison with quadratic solvers}\label{app:qubo_to_pubo}

\begin{figure}
	\centering
	\includegraphics[width=0.92\linewidth]{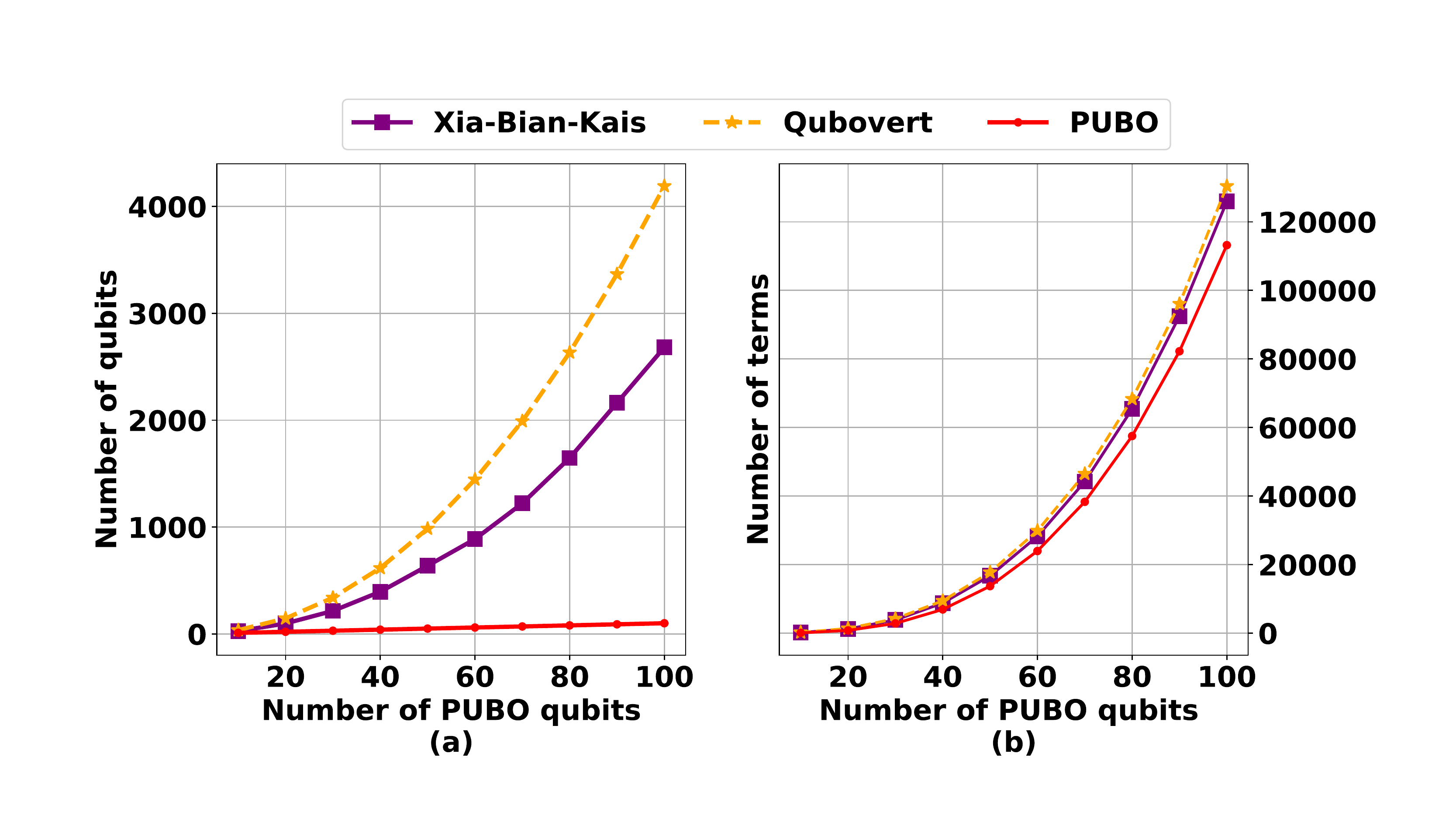}
	\caption{Comparison of the overhead of transformations with Qubovert and Xia-Bian-Kais method in locality reduction of original PUBO problems: (a) problem size (number of bits); (b) total number of terms in the energy function.}
	\label{fig:qubit_scaling2}
\end{figure}

Since all the algorithms benchmarked above are still an active area of research, it is of interest to compare results with commonly used blackbox solvers. However, to our knowledge, there exist no such solvers specifically optimized for PUBO with $k>2$. Existing solvers are limited to the quadratic domain, so, in order to use them, we had to convert our problems to the QUBO form. To this end, we used two methods: one based on the paper by Xia, Bian, Kais \cite{electronic_ising_transform} and the other from the open-source Python library Qubovert \cite{qubovert}. While the Xia-Bian-Kais method was able to produce QUBOs with fewer variables, the transformations took considerably longer to execute. 
However, for QUBOs with fewer qubits, quadratic solvers execute more rapidly, compensating for the pre-processing time. Both methods showed similar performance in terms of the minimum energies found, so here we report the results obtained with Qubovert pre-processing (Fig.~\ref{fig:qubit_scaling2}).

The solvers used for benchmarking were as follows.
\begin{enumerate}[label=\roman*]
    \item Simulated annealing, implemented with D-Wave's open source code \cite{dwave};
    \item Gurobi's mixed integer programming solver \cite{gurobi}. Gurobi utilizes a plethora of methods such as branch-and-cut, deterministic parallel, non-traditional search, heuristics, solution improvement, cutting planes, and symmetry breaking algorithms.
    \item IBM CPLEX \cite{cplex} which utilizes methods such as simplex, barrier interior point method, and second-order cone programming. 
\end{enumerate}
To match execution times with the PUBO solvers, the time limit parameter for (ii) and (iii) was set to two minutes. However the solvers did not conform to this time limit in all cases: some executions took as long as ten hours. Because of this slow performance, we collected only a single sample for each problem instance with these algorithms. For (i), since the time limit parameter setting was not available, other parameters were tuned to have the execution time close to two minutes. All QUBO solvers were executed on a 2.3 GHz Quad-Core Intel Core i7 processor.

\begin{figure*}
		\centering
		\includegraphics[width=\textwidth]{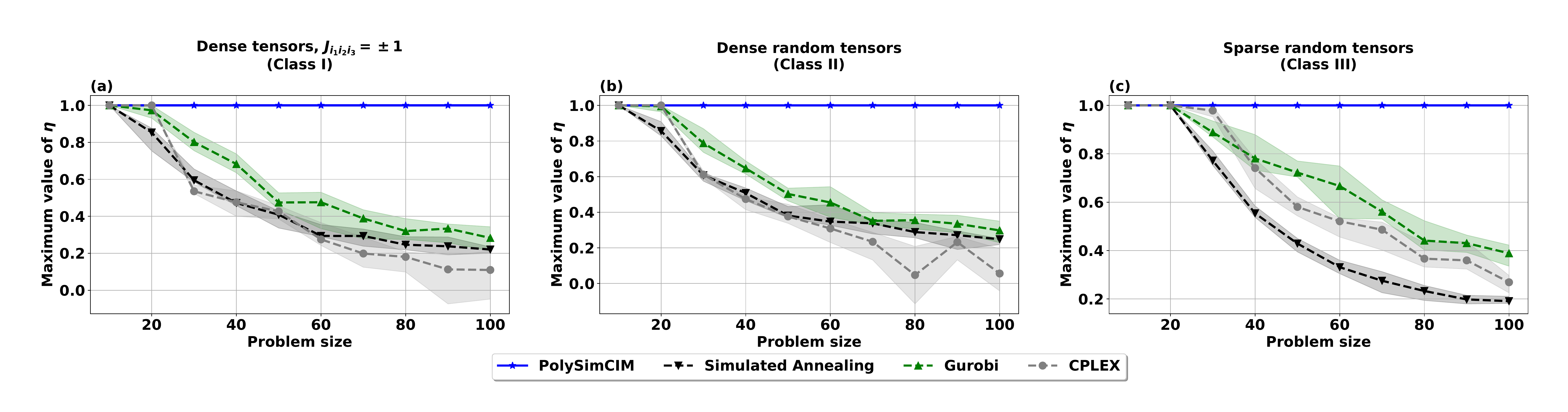}
		\caption{Performance comparison of PolySimCIM against QUBO solvers. PolySimCIM and simulated annealing produced 2000 samples per problem instance, of which the best $\eta$ value is chosen. Gurobi and CPLEX produced only one sample and were run only once per problem instance due to their slow convergence. Similar to Fig.~\ref{fig:benchmarking}, the median of these values over 10 problem instances corresponding to a specific problem class and size is shown. Shaded regions show the 25-75 percentiles of these values over each set of problem instances.}
		\label{fig:qubo_benchmarking}
	\end{figure*}

As seen in Fig.~\ref{fig:qubo_benchmarking}, the blackbox QUBO solvers performed significantly poorer than most of the PUBO solvers discussed in Sec.~\ref{sec:benchmarking}. This is likely due to the significant overhead introduced by the locality reduction scheme [Fig.~\ref{fig:qubit_scaling2}(a)].
A plausible inference from these results is that solving optimization problems in their native form generally leads to higher quality solutions. Even though the QUBO solvers we used are state-of-the-art in their own domain, they were unable to perform competitively outside it. A perhaps more speculative conclusion is that solving problems of higher than the second degree with current quantum annealers is not a advisable approach, as the latter are limited to the quadratic domain. 

	\section{Application to real-world problems}
	We now apply PolySimCIM to solve protein folding and electronic structure problems. 
	Both problems can be formulated in the PUBO framework. In previous research, they were subsequently converted to QUBO and then solved using quantum annealers, such as D-Wave 2000Q \cite{Perdomo-Ortiz2012, quantum_ising_to_classical}. Applying PolySimCIM allows us to eliminate this second conversion, and solve the problem directly as a PUBO, thereby achieving superior performance. This being said, the ``PUBO route" is not the only, and oftentimes not the optimal, method to tackle them \cite{RevModPhys.92.015003, DAVIDSHERRILL1999143, Thachuk2007, Guo2017}, as we see below.
	
	\subsection{Lattice protein folding}\label{sec:protein_folding} 
	Being initially a mere chain of amino-acid residues, soon after the synthesis each protein molecule folds into a unique spatial structure, corresponding to the minimum of free energy.
	This structure determines most of the protein behaviour, while proteins with damaged or incorrect folds are unable to function properly.
	As proteins form arguably the most versatile and ubiquitous class of biomolecules, the problem of finding the spatial configuration of a protein from its primary amino-acid sequence --- also known as \emph{the protein folding problem} --- is of paramount importance in computational biology~\cite{Dill2008, Sali1994, Dill2012, Englander2014}.
	
	Over the last half-century, this problem has been approached with a variety of methods. 
	One family of approaches considers interaction field forces between different amino-acid residues and solves equations of motion for the latter, thus recovering both the folding process and its outcome~\cite{Hansmann1999}.
	The second large family are Monte Carlo methods, where random trial moves of a protein chain are iteratively tested. 
	These moves are then either accepted or rejected based on the Boltzmann weight of the newly obtained conformation~\cite{Zhang1992, Kolinski1994}.
	Finally, techniques based on machine learning have recently come to the fore, with a deep neural network predicting the distances between amino-acid pairs and the angles between chemical bonds connecting those pairs~\cite{Noe2020, Senior2020}.
	Note that this approach relies heavily on the extensive database of known protein structures acquired over the last decades.

	\begin{figure}
		\centering
		\includegraphics[width=0.8\linewidth]{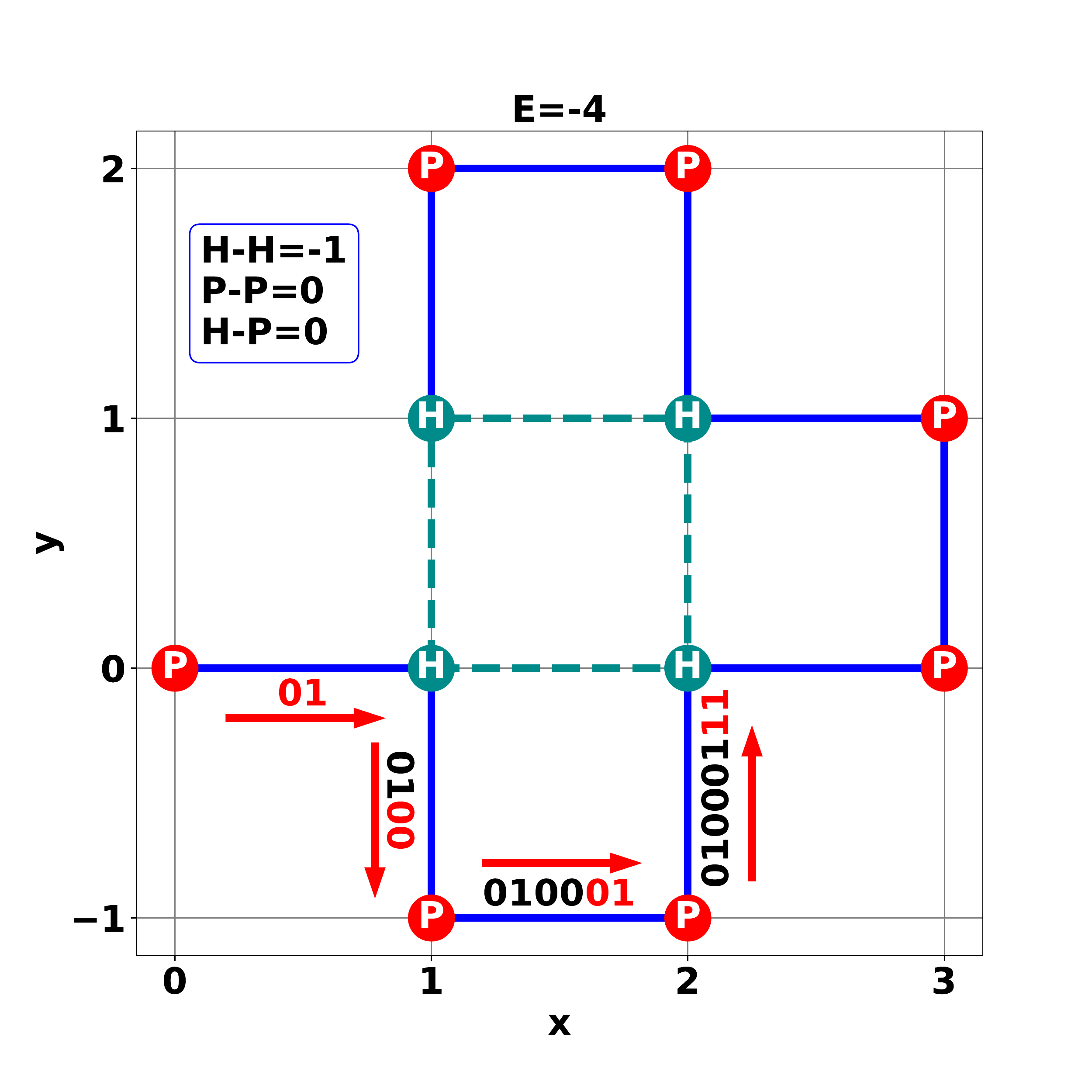}
		\caption{Lattice model of protein folding. Interactions between neighbouring amino-acids are shown as dark cyan dotted lines. Encoding of the first four turns is shown with last two red digits indicating the turn direction. The figure is specific to the PHPPHPPHPPH sequence in the PH model. }
		\label{fig:folding}
	\end{figure}

	However, despite the remarkable progress achieved so far, no ``silver bullet'' algorithm to fully predict the three-dimensional structure of a protein is yet developed. 
	To gain insight into the mechanism of protein folding, simplified models are sometimes considered, such as the 2D square lattice model (Fig.~\ref{fig:folding}). Babbush {\it et al.}~showed that such a model can be reduced to PUBO, which paves the way to applying quantum annealers and their simulators in the protein folding problem~\cite{Babbush2014}. Although lattice protein folding is a toy-model, it has been shown to predict model protein tertiary structures to remarkable accuracy~\cite{Babej2018}. 
	In what follows, we briefly discuss the idea behind this reduction.
	
	Suppose a protein consists of $M$ amino-acid residues and is embedded in a 2D square lattice.
	To fully define a planar conformation of the protein, one should specify directions of $M - 1$ turns of the protein chain.
	There are only four possible turn directions. Under the conventions of the ``Turn Ancilla" model~\cite{Babbush2014}, each turn is encoded in a pair of binary variables as follows: 01 --- ``right", 11 --- ``up", 10 --- ``left", 00 --- ``down"  (Fig.~\ref{fig:folding}).
	Hence the fold of a protein can be described with $2 (M -1)$ binary variables. 
	In fact, the first three bits can be set to arbitrary values without loss of generality, and so the fold configuration in this model is described by $2M-5$ bits. 
	The configuration energy $E_{\rm pair}(\bs)$ 
	is then made up of interaction energies between pairs of neighbouring amino-acids. 
	There exist a variety of models for evaluating these energies. 
	In a simple hydrophobic-polar (HP) model all amino-acids are classified as hydrophobic or polar, and the interaction strength depends on which of these two classes the interacting molecules belong to.
	In a more intricate Miyazawa-Jernigan (MJ) model, the energy depends on the specific types of the interacting amino-acids. Both models are detailed in Table~\ref{tab:Interactions}.
	
\begin{figure}
		\centering
		\includegraphics[width=0.8\linewidth]
		{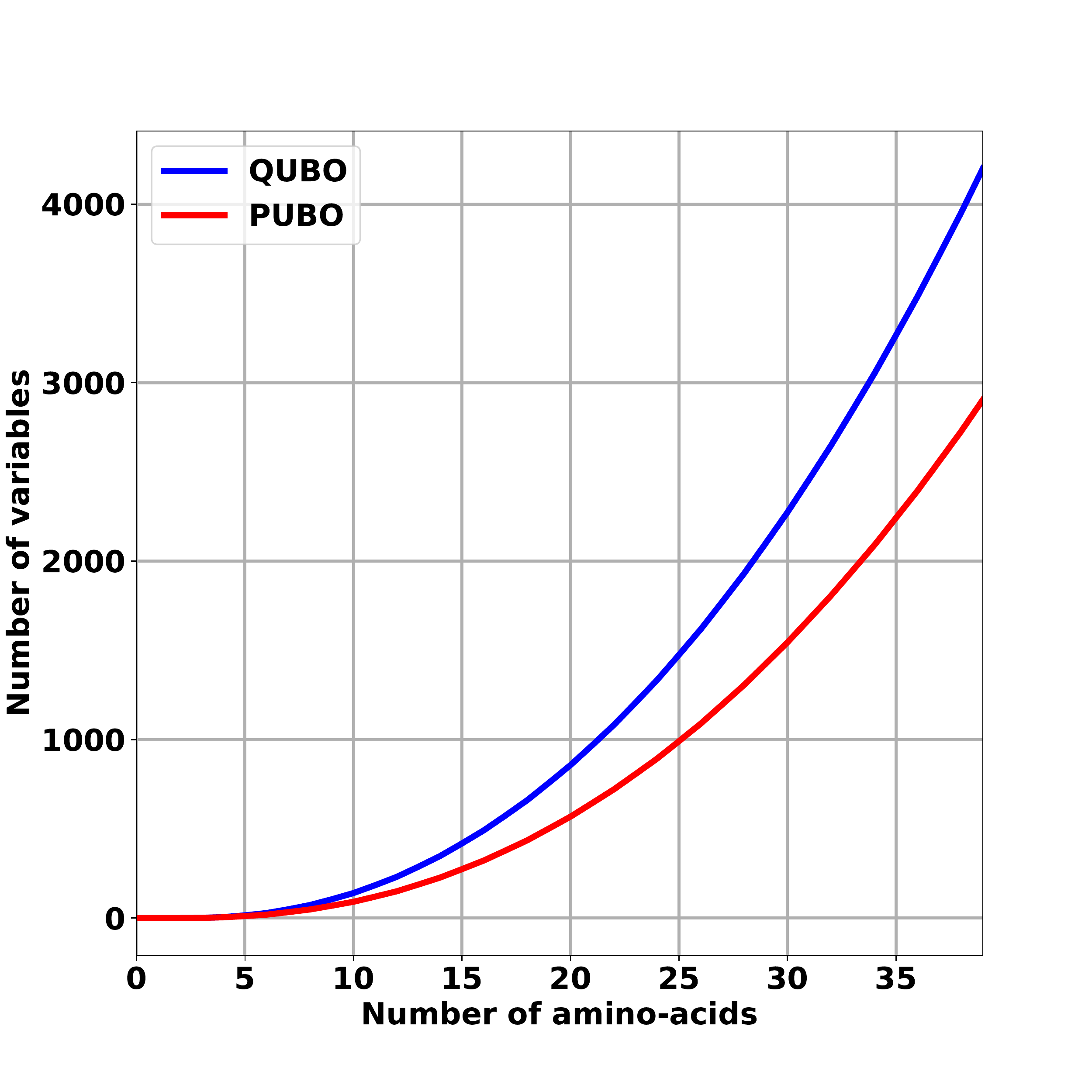}
		\caption{QUBO and PUBO problem sizes in the protein folding problem under the lattice model as a function of the amino-acid sequence length. The calculation is outlined in Appendix \ref{app:folding_nq}.}
		\label{fig:Number}
	\end{figure}

	Two further modifications are needed to express the interaction energy in the PUBO form. First, unphysical fold configurations (i.e.~those in which chain fragments overlap or intersect) are penalised by an additional energy term $E_{\rm penalty}(\bs)$.
	Second, the initial vector of $2M - 5$ variables is supplemented with ancillary variables, which are necessary to determine which amino-acids are neighbours.
	The details of these transformations are beyond the scope of our paper, but detailed in Ref.~\cite{Babbush2014}.
	The resulting energy function $E(\bs) =E_{\rm pair}(\bs) + E_{\rm penalty}(\bs)$ is then of fourth degree. 
	It can be either reduced to the second-degree (QUBO) function at the cost of adding more binary variables (Fig.~\ref{fig:Number}) or treated directly using PolySimCIM.
	
	We applied PolySimCIM to four proteins: three of lengths 6, 7 and 8 in the framework of the MJ model, and one of length 11 under the HP model. The choice of the first chain (PSVKMA) has been motivated by an existing result obtained with a D-Wave quantum annealer~\cite{Perdomo-Ortiz2012}.  In this work, the ground state of the PSVKMA sequence was found with a success probability of $0.13\%$ on the D-Wave quantum annealer using 81 qubits. The next two chains (PSVKMAP and PSVKMAPS) are obtained from the first one by sequentially appending arbitrary amino-acids. 	The largest chain (PHPPHPPHPPH) was chosen from Ref.~\cite{Rostoker2005IterativeMC}. 
	
	
	
\begin{figure*}
		\centering
		\includegraphics[width=\textwidth]{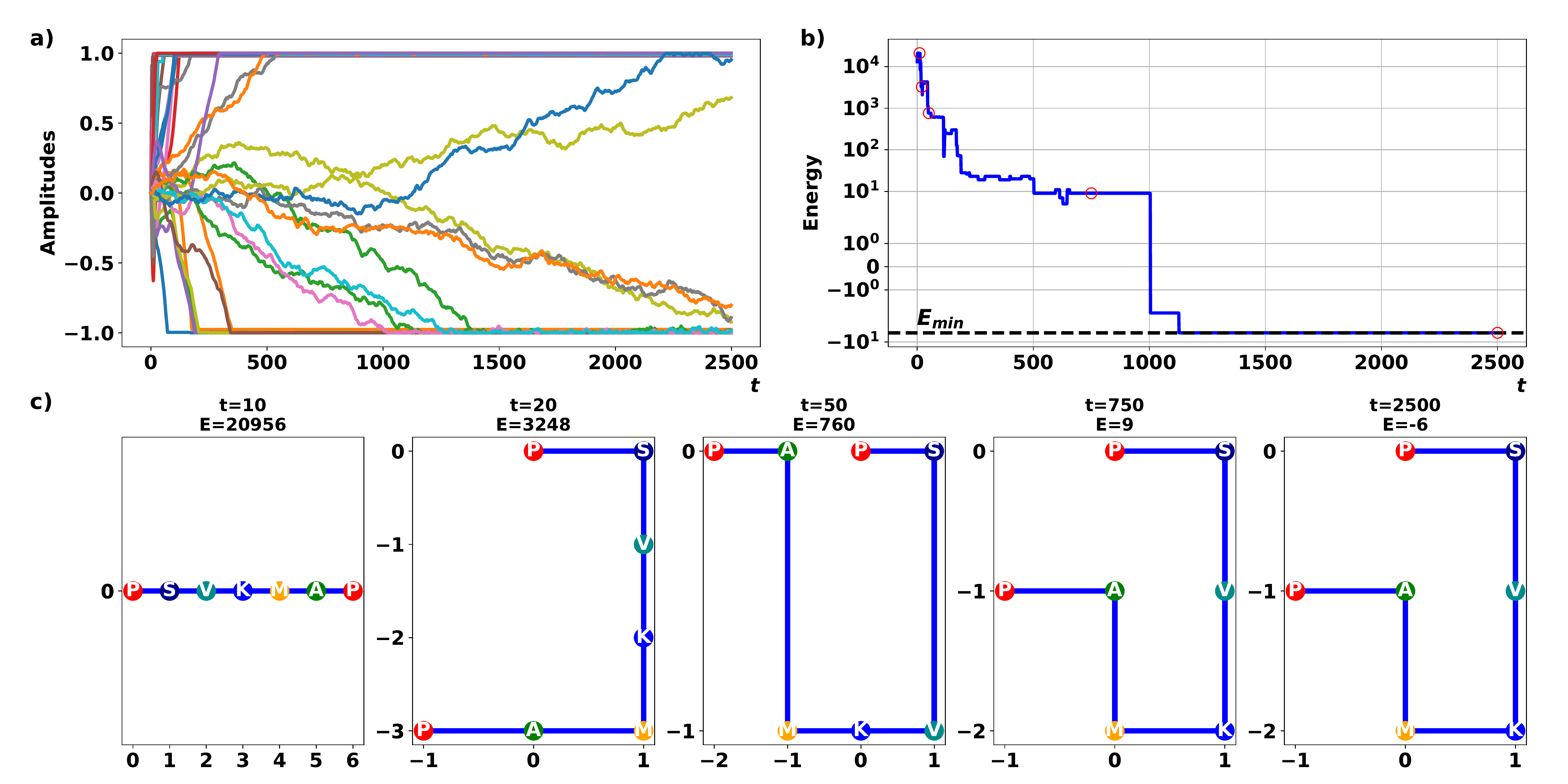}
		\caption{(a) Evolution of amplitudes in a successful run of PolySimCIM for PSVKMAP amino-acid sequence; (b) Evolution of $E(\bx)$ in the same run ($E_{\min}=-6$); (c) Protein conformations at five moments in algorithm runtime [red circles in Fig.~\ref{fig:Amplitudes}(b)].}
		\label{fig:Amplitudes}
	\end{figure*}

The results are summarized in Table  \ref{tab:Sequences}. In all four cases the fold configuration with the lowest known energy has been found successfully, which was confirmed by a brute force search. The evolution of variables in a successful run of PolySimCIM for the PSVKMAP sequence is shown in Fig.~\ref{fig:Amplitudes}(a).
	One can see that, while most variables saturate relatively quickly, a few others need a considerable number of iterations to ``commit" to a binary value.
	As a rule, the variables that take longer to stabilise are the ancillary ones, rather than the $2M-5$ variables encoding the spatial structure of the protein. This is evident in Fig.~\ref{fig:Amplitudes}(b,c): at $t=750$ the protein molecule has already folded correctly, but $E(\bs)=9$, which differs from the final optimal value $E_{\min}=-6$. In other words, the algorithm may be stopped before all variables are stabilised and still yield the optimal conformation. 

	
	\begin{table}
		\begin{center}
			\begin{tabularx}{0.9\linewidth}{ |>{\centering}X|>{\centering}X|>{\centering}X||c|c|c|c|c|c| } 
				\hline
				\multicolumn{3}{|c||}{ Hydrophobic-polar}& \multicolumn{6}{c|}{ Miyazawa-Jernigan} \\
				\hline
				HH & HP & PP & PS & PK & PA & SM & VA & VS \\
				\hline
				$-1$ & 0 & 0  & $-0.5$ & $-1$ & $-2$ & $-3$ & $-4$ & $-5$ \\
				\hline
			\end{tabularx}
			\caption{Values of interaction strengths between amino-acids for the Hydrophobic-polar and Miyazawa-Jernigan models. In the framework of the latter P, S, V, K, M, A stand for  Proline, Serine, Valine, Lysine, Methionine and Alanine, respectively.}
			\label{tab:Interactions}
		\end{center}
	\end{table}
	

	\begin{table}
		\begin{center}
			\begin{tabular}{ | m{2.5cm} | m{1cm}| m{1 cm} |m{1cm}|m{1.4 cm}| } 
				\hline
				\multicolumn{5}{|c|}{Miyazawa-Jernigan model}\\
				\hline
				Primary sequence & Energy& $N_{\rm PUBO}$& $N_{\rm QUBO}$ &$p_{\rm succ}$\\
				\hline
				PSVKMA & $-6$& 19 &28& 1.0\\
				\hline
				PSVKMAP & $-6$& 33 & 49&1.0\\
				\hline
				PSVKMAPS & $-9$& 48 &73&$10^{-4}$\\ 
				\hline
				\multicolumn{5}{|c|}{Hydrophobic-polar model}\\
				\hline
				PHPPHPPHPPH & $-4$ & 104 & 168& $2.5\cdot 10^{-5}$\\
				\hline
			\end{tabular}
			\caption{Results of PolySimCIM applied to four amino-acid sequences. 	The columns are the energy of the optimal fold, number of bits in the QUBO and PUBO settings and the success probability $p_{\rm succ}$, defined as the number of runs where the algorithm achieved the ground state energy divided by the total number of runs.  }
			\label{tab:Sequences}
		\end{center}
	\end{table}
	
	\subsection{The electronic structure problem}\label{sec:quantum_chemistry}
	
	Most of the molecule properties can be derived from its ground state wave function (also known as \emph{the electronic structure}), and thus an ability to calculate the latter --- or at least the ground state energy --- is crucial for quantum chemists.
	To find it, one usually writes down the Schr\"odinger equation for a given molecule, transforms it to a second quantised form, and then applies one of the quantum chemistry methods developed over the last half-century --- we refer the reader to a classic textbook on quantum chemistry~\cite{szabo_ostlund} for more details.
	Alas, the Schr\"odinger equation in the second quantised form is computationally tractable only for small molecules as the size of system Hilbert space grows exponentially with the number of electrons.
	
	An approach to facilitate electronic structure calculations with quantum annealers was proposed in Ref.~\cite{quantum_ising_to_classical}.
	Its first step is to rewrite the second quantized Hamiltonian in the qubit form using any of relevant transformations developed so far (e.g. Jordan-Wigner, Bravyi-Kitaev etc.)~\cite{McArdle2020}.
	The Hamiltonian then takes the form
	\begin{equation}\label{equ:qubit_Hamiltonian}
		\hat{H} = \sum_{i, \alpha}h_{\alpha}^i \hat{\sigma}_{\alpha}^i + \sum_{\substack{i,j \\ \alpha, \beta}}h_{\alpha \beta}^{ij} \hat{\sigma}_{\alpha}^i \hat{\sigma}_{\beta}^j + \sum_{\substack{i,j,k \\ \alpha, \beta, \gamma}}h_{\alpha \beta \gamma }^{ijk} \hat{\sigma}_{\alpha}^i \hat{\sigma}_{\beta}^j \hat{\sigma}_{\gamma}^k + \ldots.
	\end{equation}
	Here Latin alphabet indices enumerate the qubits and range from $1$ to $M$, while Greek letters point which of Pauli operators $\left\{ I, X, Y, Z \right\}$ constitute the term.

	The Hamiltonian \eqref{equ:qubit_Hamiltonian}  includes all three Pauli operators and is hence inherently quantum. 
	Ref.~\cite{quantum_ising_to_classical} makes it compatible with the classical PUBO problem by mapping it onto a Hamiltonian 
	\begin{equation}\label{equ:mapped_Hamiltonian}
		H = \sum_{\mathcal{S} \subset\{1,\ldots,rM\}}h_{\mathcal{S}}\prod_{i \in \mathcal{S}}s^{i},
	\end{equation}
	 which is a function of $rM$ bits that can take on values $s\in\{-1,1\}$. 
	Here $r$ is some integer number which allows fine-tuning the method accuracy.
	The details of this mapping are outside the scope of this paper, but can be found in Ref.~\cite{quantum_ising_to_classical}. 
	Note that the case of $r=1$  corresponds to the Hartree-Fock solution.
	
	To employ quantum annealers, Xia~\emph{et~al.}~\cite{quantum_ising_to_classical} propose to subsequently reduce this Hamiltonian to a 2-local (QUBO) form. This reduction however introduces a dramatic scaling overhead.
	The authors estimate the number of terms in Hamiltonian~\eqref{equ:mapped_Hamiltonian} to scale as $ \mathcal{O}\left(2^M r^2 M^4\right)$ and the number of bits in the final QUBO Hamiltonian as $\mathcal{O}\left(2^M r^2 M^7\right)$. 
 	Streif {\it et al.} report that due to such prohibitive scaling only very small molecules can actually be addressed with quantum annealers \cite{dwave_report}.
	In particular, the authors found the ground state of H$_2$ ($M=2$) with the D-Wave 2000Q quantum annealer, requiring the scaling factor value as high as $r=16$ to obtain a good agreement with exact quantum calculation. 
	However, the full treatment of a somewhat more complex molecule LiH ($M=10$) proved to be impossible on this machine as the final QUBO Hamiltonian required more qubits than was available in D-Wave 2000Q.
	To embed this problem onto this machine, the authors resorted to restricted active space methods and limited the number of orbitals which can be occupied by electrons, but then they were not able to obtain states with energies lower than those given by the Hartree-Fock method \cite{dwave_report}.  
	\begin{figure}
		\centering
		\includegraphics[width=0.92\linewidth]{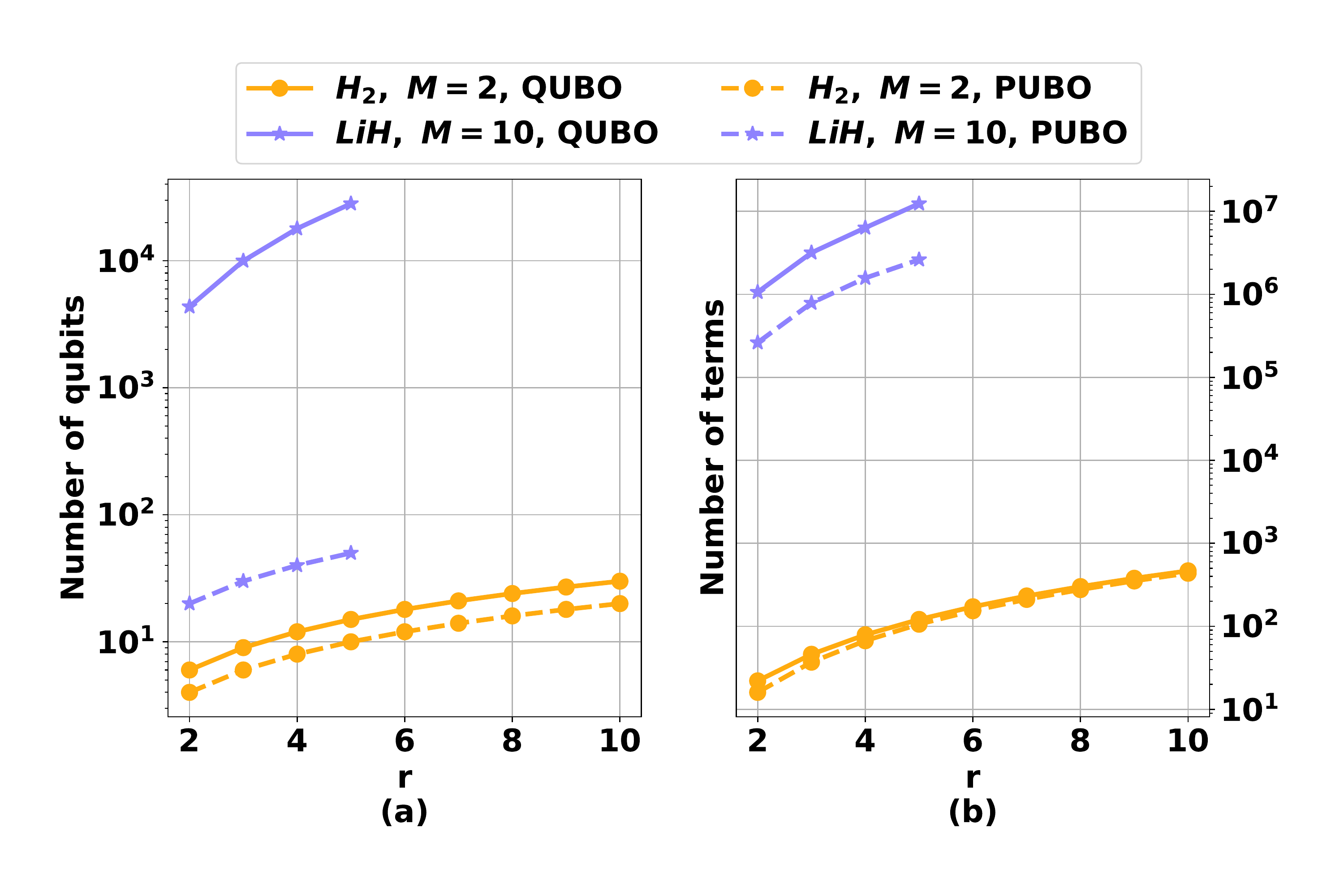}
		\caption{Comparison of QUBO and PUBO approaches to electronic structure calculations: (a) problem size (number of bits); (b) number of terms in the Hamiltonians obtained according to~\cite{quantum_ising_to_classical}. The calculation for LiH is limited to $r\le 5$ because higher values of $r$ required prohibitively long compute time.}
		\label{fig:qubit_scaling_es}
	\end{figure}
	
	\begin{figure*}
		\centering
		\includegraphics[width=0.92\linewidth]{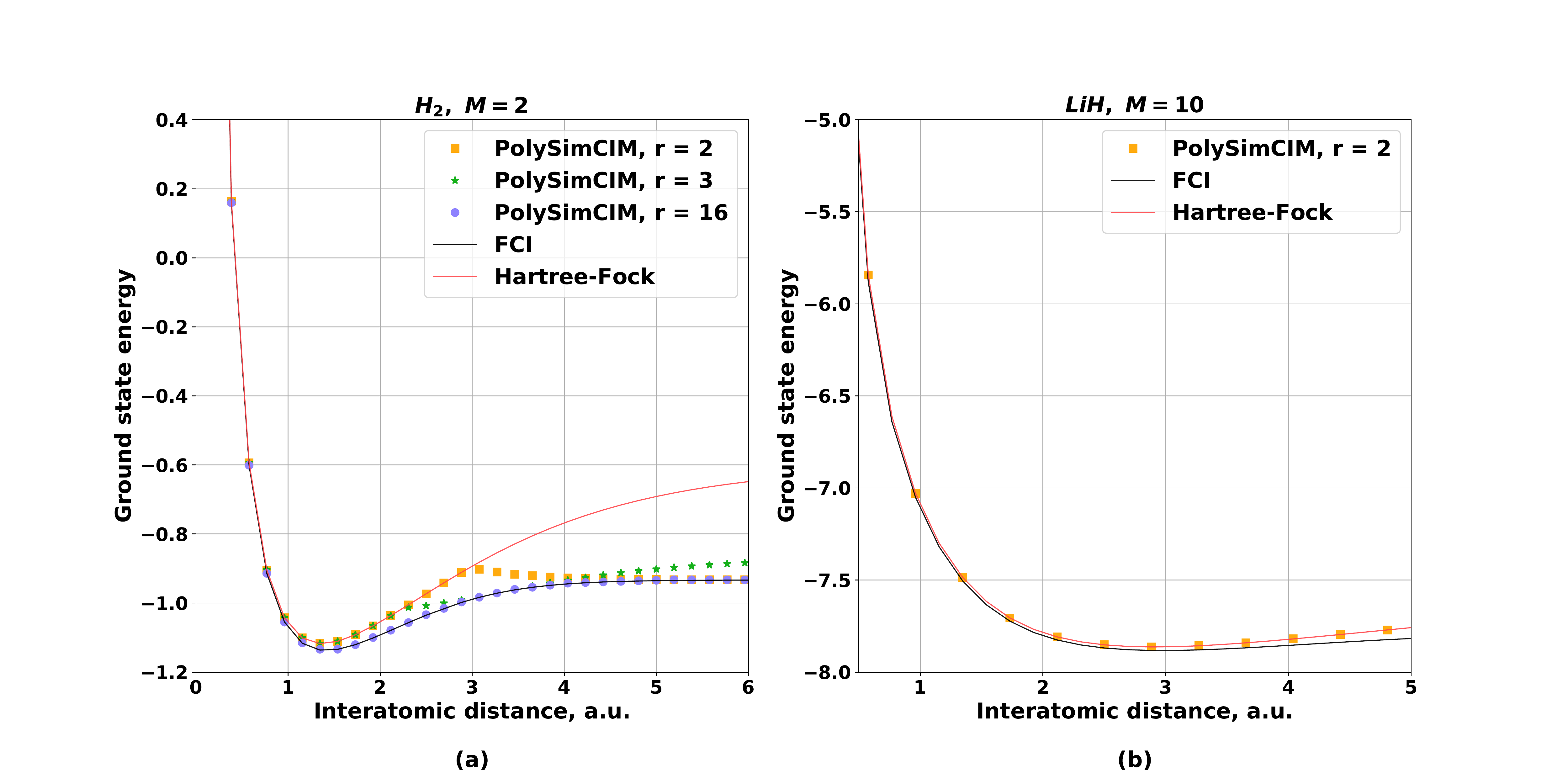}
		\caption{a) Potential energy surfaces restored by PolySimCIM for (a) H$_2$, $r \in \{2, 3, 16\}$; (b) LiH, $r= 2$.  Solid black curves represent exact potential energy surfaces, solid red curves correspond to the Hartree-Fock solution. FCI stands for Full Configuration Interaction \cite{szabo_ostlund}, a quantum chemistry method, which enables exact determination of the ground state energy.}
		\label{fig:pubo_simcim}
	\end{figure*}
	
	These circumstances motivated us to apply PolySimCIM directly to the Hamiltonian \eqref{equ:mapped_Hamiltonian} in order to avoid exponential scaling of the  number of bits. 	We attempted PUBO optimization for H$_2$ and LiH molecules. 
	These molecules are linear, and we optimised the ground state energy as a function of the internuclear distance. 
	We used the quantum chemistry package \emph{Psi4}~\cite{psi4} to calculate two- and four-body integrals in the fermionic molecular Hamiltonians. 
	We also used a software package \emph{OpenFermion}~\cite{open_fermion} and applied symmetry-conserving Bravyi-Kitaev transformation~\cite{sk_bravyi_kitaev} to both molecules so that the resulting Hamiltonians~\eqref{equ:qubit_Hamiltonian} had 2 and 10 qubits correspondingly.
	
	Figure \ref{fig:qubit_scaling_es}(a) shows the characteristic numbers of bits in the PUBO and QUBO Hamiltonians for the two molecules. 	For H$_2$, these numbers are different by only $1.5$, but for LiH this gap reaches several orders of magnitude. While this comparison appears favourable for PUBO,  the problem complexity depends not only on the number of bits, but also on the number of terms in the Hamiltonian. These numbers are comparable for the two cases and scale exponentially with $r$ [Fig.~\ref{fig:qubit_scaling_es}(b)].  Even for $r=2$, the PUBO and QUBO Hamiltonians have, respectively, $\sim2\cdot 10^5$ and $\sim 10^6$ terms, compared to 631 in the initial molecular Hamiltonian \eqref{equ:qubit_Hamiltonian}. This scaling arises because, unlike the problems studied in previous sections, the degree of terms in the PUBO Hamiltonian \eqref{equ:mapped_Hamiltonian} is limited only by the number of bits. As a result, while the problem still appears simpler in the PUBO formulation, it remains computationally expensive to solve.

	
	The results are presented in Fig.~\ref{fig:pubo_simcim}. The calculations for H$_2$ have been performed for $r \in \{2, 3, 16\}$, while for LiH we considered only $r = 2$ as higher values of $r$ would result in prohibitive computation overheads.
	For each setting, PolySimCIM made 500 independent runs of 2000 update steps.
	The same hyperparameters were used for all internuclear distances. In all cases studied, PolySimCIM produced solutions equal to the ``brute force'' ones obtained by exact \red{optimization} of the corresponding PUBO Hamiltonians~\eqref{equ:mapped_Hamiltonian}; it was possible to calculate the latter as the  Hamiltonians had at most 32 bits. 
	
	However, these solutions were not always consistent with the ground states of the quantum Hamiltonian  \eqref{equ:qubit_Hamiltonian}.
	For the H$_2$ molecule, increasing the value of $r$ leads to consistent improvement of the approximation. For $r=16$, the computed energy lies within the chemical accuracy from the surface obtained by direct diagonalization of the Hamiltonian  \eqref{equ:qubit_Hamiltonian}. For LiH, the calculation with $r=2$ did not provide better solution than Hartree-Fock.
	Thus, even though PolySimCIM allowed us to perform calculations for system sizes that were  previously unachievable using quantum annealers, the approximating the quantum molecular Hamiltonian by a PUBO Hamiltonian does not  appear to be a viable approach for electronic structure calculations. The reasons are (i) exponential scaling of the number of terms in the PUBO Hamiltonian and (ii) high values of $r$ required to obtain the desired accuracy. 
	
	\section{Discussion}\label{sec:conclusion}
	
	We have extended the  quantum-inspired SimCIM algorithm to cover polynomial unconstrained binary optimization with degrees $k>2$. We  benchmarked PolySimCIM on a variety of PUBO graphs in comparison with (1) other QUBO algorithms modified to solve PUBO and (2) blackbox QUBO algorithms, for which the PUBO problems were converted to the QUBO format. In both cases, PolySimCIM proved superior.
	
	We have also applied PolySimCIM to the protein folding and electronic structure problems, surpassing the performance of quantum annealers and their simulators. However, despite this success, large-scale electronic structure calculations are unlikely to succeed with this approach. The fault appears to be not in the algorithm itself, but rather in mapping this problem onto PUBO: the resulting energy function scales exponentially with the molecule size and quickly becomes intractable. This reinforces our hunch from Sec.~\ref{app:qubo_to_pubo} that optimization problems are best solved in their original format. 
	
	The main conclusions of this work are that (1) PolySimCIM is a state-of-the-art algorithm for PUBO problems, and (2) for not ``native PUBO" combinatorial or quantum optimization problems,  PolySimCIM provides a good testbed to find out whether conversion to PUBO is the optimal route towards solution.
	
	We suppose that the performance of PolySimCIM can be improved by adding time-dependent correction to amplitude inhomogeneity akin to Ref.~\cite{Goto2021}. This modification can be subject of future investigations.
	
	The extension from QUBO to PUBO is also possible in hardware. For example, in the optoelectronic implementation of a fiber optic coherent Ising machine \cite{Honjo2021, Mcmahon2016}, the gradient of pseudo-Boolean energy function is computed on a FPGA by multiplying the matrix $J_{i_{1}}i_2^{(2)}$ by a vector of measured quadratures. This operation can be extended to PUBO by multiplying the PUBO tensor $J_{i_{1}\ldots i_{k}}^{(k)}x_{i_{2}}$ by multiple copies of the measured quadrature vector. Moreover, PUBO can be implemented by pure analogue means, without any digital computing. Aside from the aforementioned Ref.~\cite{Stroev2021}, in which an implementation via polariton condensates is proposed, a high-order coherent Ising machine can be based on the frequency conversion in a nonlinear crystal \cite{Kumar2020}. Such an experimental setup can solve polynomial pseudo-Boolean functions with all-to-all connections.
	
	
	\vspace{1em}
	\acknowledgements
	This work was supported by Russian Science Foundation (19-71-10092).
	The part on quantum chemistry is also partially supported by Nissan Research (QUBO analysis). D.A. acknowledges Nikita Stroev for fruitful discussions and help with TGD+CC algorithm realisation.
	A.F. is supported by the Priority 2030 program at the National University of Science and Technology ``MISIS'' under the project K1- 2022-027 (analysis of the method).
	
\appendix
\section{Number of bits in the protein folding problem.}\label{app:folding_nq}
We briefly recap the method for calculating the number of bits in a protein molecule consisting of $M$ interacting amino acids \cite{Babbush2014}.
Three terms $n_{\rm phys}$, $n_{\rm penalty}$, $n_{\rm pair}$ and $n_{\rm reduction}$ contribute to the total number of bits in the PUBO model. 
The first term corresponds to the number of bits $n_{\rm phys}$ encoding the spatial configuration of the protein. 
The second term $n_{\rm penalty}$ emerges from the penalty part of the pseudo-Boolean function designed to avoid physically impossible conformations. 
The third term $n_{\rm pair}$ is the number of ancilla variables, which is equal to a number of potential interactions between amino acids in the sequence. 
Let us note that the interactions between amino acids are possible only if the difference between their positions in the primary sequence is greater than 3. It means that first and third aminoacids cannot interact, but the first and fourth can. When the PUBO pseudo-boolean function is reduced to the QUBO one, an additional term $n_{\rm reduction}$ is added to the number of bits:
\begin{equation}
	\begin{gathered}
		N_{\rm PUBO}=n_{\rm phys}+n_{\rm penalty}+n_{\rm pair};\\
		N_{\rm QUBO}=n_{\rm PUBO}+n_{\rm reduction}.
	\end{gathered}
\end{equation}
This reduction can be realised in a variety of ways, and choosing the most efficient procedure is an NP-hard problem \cite{BOROS2002155}. In this manuscript, we use the method of Ref.~\cite{Babbush2014} due to its high efficiency in the case of the ``Turn Ancilla" model. The explicit formulae for calculating the number of bits are the following:
\begin{equation}
	\begin{gathered}
		n_{\rm phys}=2M-5;\\
		n_{\rm penalty}=\sum_{i=4}^{M-4}\sum_{j=i+4}^{M}\ceil{\log_{2}(i-j)^2}[(1+i-j)\mod2];\\
		n_{\rm pair}=\sum_{i=1}^{M-3}\sum_{j=i+3}^{M}[(i-j)\mod2];\\
		n_{\rm reduction}=\sum_{i=1}^{2M-7}\sum_{j=i+2}^{2M-5}[(i-j+1)\mod2].
	\end{gathered}
\end{equation}

\section{Hyperparameter Search}\label{app:hyperparameter_search}
To find optimal hyperparameters for the algorithms, we use the M-LOOP machine learning package \cite{Wigley2016}. Within M-LOOP, we use the ``differential evolution" method for sampling the data set to train the neural network and the ``neural net" controller for the optimization itself.
The cost function has the following form of weighted sum:
\begin{equation*}
    \mathrm{cost} = 0.25\min{\left[\frac{\langle H(\bs)\rangle}{\left|\langle H_0(\bs)\rangle\right|}, 1\right]} + \min\left[\frac{\min_{N} H(\bs)}{\left|\min_{N} H_0(\bs)\right|}, 1\right]
\end{equation*}
Here $\{H(\bs)\}$ is the set of values of the energy function obtained from all runs of the algorithm. The first term corresponds to the ratio between the mean value of the energy function over all runs of the algorithm in each M-LOOP step and the modulus of the mean value in the first step. The second term corresponds to the ratio between the minimum (best) obtained value of the energy function over all runs and the modulus of the minimum value in the first step of M-LOOP. The cost function is bounded by the value 1.25 from above. This normalization allows us to improve the training of the M-LOOP neural networks and to speed up the process of finding the optimal hyperparameters. 1500 steps of M-LOOP were realized for each algorithm. The parameters are searched on the logarithmic grid in the interval $[10^{-8}, 10^{3}]$ with the exception for $\sigma$, which was searched in the interval $[10^{-8}, 10^{-0.1}]$.

	\bibliographystyle{apsrev4-2} 
	\bibliography{bibliography, PUBO_ref}
\end{document}